\documentclass[twocolumn]{article}
\usepackage{moreverb,url}

\usepackage[colorlinks,bookmarksopen,bookmarksnumbered,citecolor=red,urlcolor=red]{hyperref}

\newcommand\BibTeX{{\rmfamily B\kern-.05em \textsc{i\kern-.025em b}\kern-.08em
T\kern-.1667em\lower.7ex\hbox{E}\kern-.125emX}}

\usepackage[utf8]{inputenc}
\usepackage{subcaption}
\usepackage{graphicx}
\usepackage{booktabs}
\usepackage[numbers]{natbib}
\usepackage{xcolor}

\def\hb{\hbox to 11.5 cm{}}
\providecommand{\keywords}[1]
{
  \small	
  \textbf{\textit{Keywords---}} #1
}
\usepackage{authblk}

\title{Virtual Laboratories: Domain-agnostic workflows for research}
\date{}
\begin{document}




\author[1]{Carlos Sevilla-Salcedo}
\author[1]{Armi Tiihonen}
\author[1]{Mahsa Asadi }
\author[3]{Kevin Sebastian Luck}
\author[2]{Aras Umut Erarslan}
\author[2]{Arto Klami}
\author[1]{Samuel Kaski}
\affil[1]{Department of Computer Science, Aalto University, Espoo, Finland}
\affil[2]{Department of Computer Science, University of Helsinki, Helsinki, Finland}
\affil[3]{Faculty of Science, Vrije Universiteit Amsterdam, Amsterdam, Netherlands}



\maketitle

\begin{abstract}

    Many scientific disciplines have traditionally advanced by iterating over hypotheses using labor-intensive trial-and-error, which is a slow and expensive process. Recent advances in computing, digitalization, and machine learning have introduced tools that promise to make scientific research faster by assisting in this iterative process. However, these advances are scattered across disciplines and only loosely connected, with specific computational methods being primarily developed for narrow domain-specific applications. 
    Virtual Laboratories are being proposed as a unified formulation 
    to help researchers navigate this increasingly digital landscape using common AI technologies. While conceptually promising, VLs are not yet widely adopted in practice, and concrete implementations remain limited.
    This paper explains how the Virtual Laboratory concept can be implemented in practice by introducing the modular software library 
    VAILabs, designed to support scientific discovery. VAILabs provides a flexible workbench and toolbox for a broad range of scientific domains. We outline the design principles and demonstrate a proof-of-concept by mapping three concrete research tasks from differing fields as virtual laboratory workflows. 
\end{abstract}



\section{Introduction}
        Scientific research is often a slow process, requiring repeated examination of alternative hypotheses in search of new knowledge. In most cases, this is labor-intensive and costly, for instance, due to slow laboratory experimentation and complex processes.
        Digital tools offer transformative potential for increasing research efficiency. Among them are computational simulations and machine learning proxies, capable of predicting outcomes without manual experimentation \cite{wang2022inverse}. Others include co-researchers based on large language models (LLMs), which can act as discussion partners or perform sub-tasks on behalf of the researcher \cite{schmidgall2025agentlab,swanson2024virtual}.
        For example, in drug discovery, computational simulations have long provided valuable preliminary insights \cite{sliwoski2014computational}, generative Artificial Intelligence (AI) methods are used to propose novel candidates \cite{blaschke2020reinvent}, and LLMs are revolutionizing the ability to use historical information stored in scientific publications \cite{li2025scilit}. Nevertheless, the experimental validation of a drug candidate's efficiency remains essential. That is, the computational tools \emph{complement} rather than replace physical experiments, offering partial information that can guide subsequent testing steps. These tools have been integrated into individual research steps or short sequences of steps across various experimental science domains. However, comprehensive and unified strategies for designing entire workflows are still taking shape. Such strategies must be efficient, avoiding the need to involve a prohibitively large number of laboratory scientists in redesigning the workflow process every time the research focus shifts. As we integrate these digital tools, a key challenge indeed emerges: we need robust methodologies to help researchers understand when and how these tools should be applied to maximize their potential impact.

        Scientific inquiry is dynamic and constantly evolving, unlike static production workflows. This demands new digital-physical frameworks that can adapt to changing questions and experimental conditions. AI is poised to play a central role in this shift. It helps optimize processes, analyze vast datasets, and suggest innovative directions for exploration \cite{bobbili2023artificial, johnson2016malmo, buchanan1994role}. With the advent of powerful LLMs and reasoning engines, AI is increasingly used not only for data analysis but also for producing and interpreting natural language resources \cite{jiang2025applications} and other forms of digital information \cite{birhane2023science}. AI could even be driving the discovery cycle itself \cite{swanson2024virtual,sparkes2010towards}. A key challenge is to combine these digital tools with traditional physical measurements in ways that enhance scientific exploration rather than fragment it \cite{kluge2014combining}.

        Within this evolving landscape, the field often referred to as \emph{AI4Science}, particularly in the natural sciences, there are two major streams of progress. The first focuses on developing advanced models for specific scientific applications, such as predicting protein structures from amino acid sequences \cite{jumper2021alphafold}, generating novel molecules with desired properties \cite{loeffler2024reinvent}, or predicting new functional materials \cite{merchant2023scaling}. The second stream aims at supporting researchers more directly, for example, through algorithm recommendations for sequential experimentation \cite{wang2023booverview}, the use of LLMs to assist in literature review and other routine tasks \cite{agarwal2025litllm}, or tools that aid hypothesis generation \cite{wang2023scientific,tiukova2024genesis}. Our work relates to both streams, but does not introduce new models or tools. Instead, we focus on the research practice itself: developing frameworks that help researchers effectively use these advances as integrated parts of the scientific process rather than isolated technological achievements. We assume that various computational models, including domain-specific simulators, machine learning proxies, and AI-assisted research tools, are available and continue to evolve. Our priority is to ensure these tools are accessible across scientific domains and genuinely useful to researchers.

        Virtual Laboratories (VLs) act as digital counterparts to their physical analogues, coined after the \emph{virtual machines} in computing \cite{shirvaikar2007virtual}: They are digital environments for virtualization or emulation of the operations carried out in a physical laboratory \cite{chiu1999benefits,klami2024virtual}. The aim is to enable virtualization of arbitrary research activities without limitations, e.g., on the research domain. Within the virtual environment, the researcher can access various computational solutions for accelerated experimentation while connecting to the real world with suitable interfaces, such as conducting a physical experiment and passing the results to the VL. Note that the word virtual does not refer in this case to virtual reality (VR); it is possible to use VR solutions as interface components for a digital laboratory environment \cite{hernandez2019virtual}, but the core focus of the VL concept is in running and assisting the research operations, agnostic of the interface. In this sense, VLs link closely to the development and running of semi- or fully autonomous laboratories - physical laboratories with a high level of automation \cite{stach2021autonomous}.
        

        AI already has a prominent role in numerous scientific domains.
        In chemistry, AI has been employed to guide multi-step chemical synthesis planning \cite{schwaller2021ibm,steiner2019organic}, in biomedicine 
        AlphaFold can predict protein structures with unprecedented accuracy \cite{jumper2021alphafold}, and LLMs are being used to assist in literature review, clinical trial design, and even hypothesis generation \cite{rao2023evaluating,singhal2023large}. In a recent example, Swanson et al. \cite{swanson2024virtual} validated SARS-CoV-2 nanobodies using a constellation of an LLM investigator and a collection of complementary AI agents. In physics, symbolic regression methods like AI Feynman have uncovered interpretable physical laws directly from data \cite{udrescu2020ai}. However, many of these tools remain specifically designed for the needs of individual fields, limiting their reusability and generalizability. To broaden their impact, we advocate for the development of general-purpose AI systems designed to support the scientific discovery process itself, transcending disciplinary boundaries and minimizing the need for domain-specific reconfiguration \cite{gajos2022people}. 
        Framing research activities in terms of VLs offers such a unifying abstraction across disciplines. As more fields adopt this paradigm, it will steer the AI community toward addressing common challenges, such as hypothesis generation, experimental design, and result interpretation, that cut across scientific domains. Realizing this vision, however, will require robust and operational tools for managing and deploying VLs tools that are still largely underdeveloped today.

        \begin{figure}[t]
            \centering
            \includegraphics[width=0.9\linewidth]{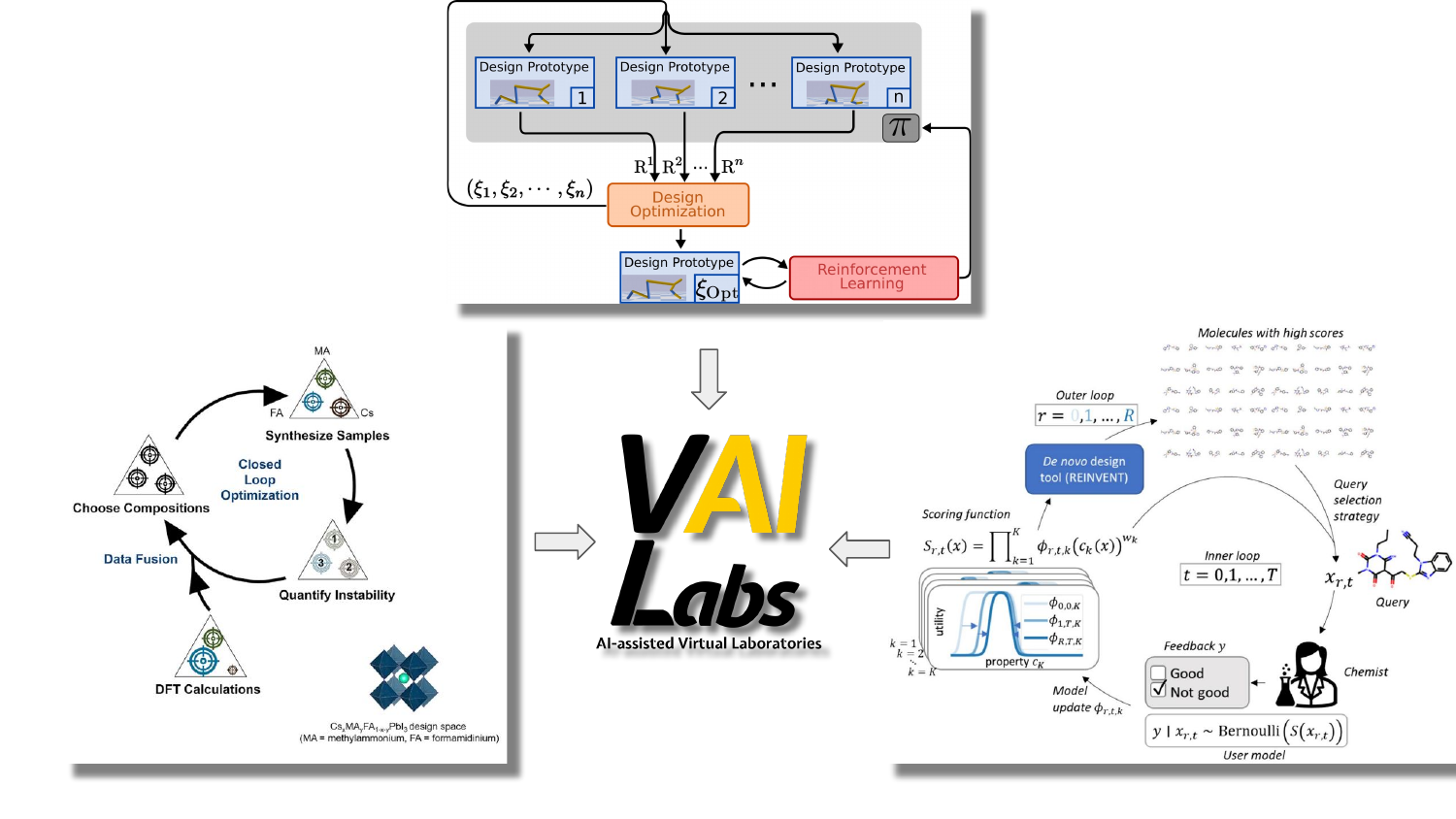}
            \caption{Mapping of experimental design use cases to VAILabs. }
            \label{fig:intro}
        \end{figure}
        
        Besides providing an overview of the current state of software support for AI-assisted research, the main goal of this paper is to provide a proof-of-concept of a software environment, VAILabs, that facilitates the operation of VLs in various scientific disciplines, as an open source, modular and domain-agnostic toolbox following the recommendations of \cite{klami2024virtual}.  
        A core element is a way of describing iterative research workflows in a systematic manner, as processes where the researcher interacts with various tools (simulations, machine learning models, digital twins, and empirical experiments), and the functionality required for running the workflow. 
        General computational tools and domain-specific models are not considered parts of the core environment but instead are accessed via general interfaces.
        
        To showcase the generality of the approach, we map three recent and diverse real-world research projects, depicted in Figure~\ref{fig:intro}, as VL workflows, capable of replicating the research but also exposing the re-occurring components for the AI community as tasks in need of improved methods.  
        The first example is a prototypical natural science optimization loop that still needs to interact with the physical world \cite{sun2021data}. The second shows how recent advanced reinforcement learning algorithms that need to optimize both the robot's body shape and type \cite{luck2020data}, now in a fully digital fashion, can be run on the same platform. Finally, the last example is an instance of the massively important scenario that involves active human decisions in the form of active learning implementation with human in the loop \cite{sundin2022human}.
        Through these examples in heterogeneous domains, we demonstrated the feasibility of the approach across different scientific fields, including its use in exploration-driven laboratories with rapidly evolving experiment pipelines. 

\section{Digital Transformation in Laboratory Research}

    To provide context for the work, we outline the past work on digitalization and automation of laboratory research and education. We do not aim for an exhaustive literature review, but instead, point out examples specifically focusing on the intersection of the digital and physical worlds.

    \subsection{Virtual environments for education}

    The term \emph{virtual laboratory} was perhaps first used in the late 1990s for software applications designed to simulate real-world scientific experiments and educational activities \cite{chiu1999benefits}, as a means of providing the users with an immersive, interactive environment in which they can perform a wide range of experiments and observations without the need for physical equipment or laboratories. These digital simulations aim to replicate some of the physical, chemical, biological, or mathematical phenomena encountered in traditional laboratories. However, the connection to the physical world is here completely omitted, and the main use remains in educational contexts, e.g. \cite{abouelenein2024impact,may2023online,sapriati2023effect,stauffer2018labster,achuthan2011value,yaron2010chemcollective,perkins2006phet,safaric2001telerobotics}. The users can simulate the research process and practice different stages, but the goal is not to make new scientific discoveries.
    The role of these environments is best understood through concrete examples:
    \begin{enumerate}
        \item PhET Interactive Simulations \cite{perkins2006phet} offers a vast collection of free interactive maths and science simulations. Their simulations cover physics, chemistry, biology, and more.
        \item ChemCollective \cite{yaron2010chemcollective} is designed for chemistry students, allowing users to explore various experiments and chemical reactions. 
        \item Labster \cite{stauffer2018labster} provides a variety of VL experiences for higher education institutions, covering topics such as biology, chemistry, and physics.
        \item Virtual Labs by Amrita University \cite{achuthan2011value} offers an extensive collection of interactive simulations for engineering and science disciplines. 
    \end{enumerate}

    \subsection{Laboratory automation}

    Starting from the opposite perspective, laboratory automation has progressed towards complete automation of empirical research. First systems capable of completely autonomous operations started appearing in the early 2000s \cite{dittrich2006lab,king2004functional}, formalized with the concept of 'robot scientist' \cite{sparkes2010towards} that carries out independent scientific research by algorithmic execution of the whole research loop, from hypothesis generation to experimentation and conclusions. The early examples were highly limited in scope, but already capable of making contributions to scientific knowledge \cite{williams2015cheaper}. Despite significant progress in the capability of robot scientists, the best examples are still for narrowly defined tasks: At the core of the loop is a specific laboratory automation process, designed for executing a specific type of laboratory work. Autonomous, semi-autonomous, or self-driving laboratories \cite{hippalgaonkar2023knowledge, maffettone2023missing} are all aiming for highly automated laboratory workflows capable of planning, experimentation, and analyses. There are already a few motivating examples on such workflows in materials discovery \cite{szymanski2023autonomous, merchant2023scaling} but the results have also raised a discussion on the best ways to ensure the reliability of the results \cite{leeman2024challenges}, which also motivates the need for VLs to exhaustively test such processes. While important progress has been made on more general concepts, such as automatic hypothesis generation and AI for driving the experimental loop in this context \cite{tiukova2024genesis}, instantiating the robot scientist for any new discipline requires significant effort in automation, and research in areas where experimentation cannot be easily automated (e.g. social sciences) is completely out of scope.

    \subsection{Combining physical and digital world}

    Our work lies between the physical and the digital world. The concept of a VL in this work refers to a collection of computational tools and resources that support the entire scientific process, with an interface playing a crucial role in the laboratory's functionality.
    The laboratory is not assumed to be fully automated, but instead, we aim to complement the actual physical working environment of the researchers with the virtual counterpart.

    Several works, including \cite{klami2024virtual, abouelenein2024impact, may2023online, achuthan2021impact}, discuss the challenges in realizing the vision of high-level VLs, emphasizing critical issues that continue to limit their widespread adoption. A key barrier is the technical complexity involved in developing advanced computational models and simulations that can faithfully replicate complex scientific phenomena, an effort that demands both domain expertise and substantial resources. Another major obstacle lies in the integration of VLs with real-world data and instruments, which is not only technically demanding but also requires ongoing infrastructure maintenance. Additionally, the financial cost of creating and sustaining high-level VLs remains prohibitive for many institutions. Even when such platforms are technically feasible, ensuring a high-quality user experience—one that is intuitive and engaging enough to encourage adoption, especially for simulating complex lab procedures—poses further design challenges. Finally, scalability is an enduring issue, as VLs must often support large numbers of concurrent users without degrading performance. These factors together explain why the number of operational high-level VLs remains limited today.
    
    
    However, the development and adoption of VLs would be underpinned by an array of advantages, which makes them worth pursuing. Digitalization of the laboratory environment is strictly necessary to reap the cost benefits of digital experimentation, and with the exception of manual LLM support, all AI integration requires digitalization. There is an obvious cost in turning a physical laboratory into a virtual one, and the main goal of VL frameworks is to minimize it. Furthermore, providing common interfaces and software tools to reduce the need for field-specific development, and a unified platform makes it easier to integrate digital tools, from simulations and machine learning proxies to LLMs. For example, many tools for data collection, analysis, and visualization are reusable across domains and laboratories.  
    Broader advantages of VLs in re-shaping the focus of AI scientists are outlined in Section \ref{sec:VAIlab}.
    



    \subsection{Workflow managers}

        As a crude approximation, iterative research can be seen as a sequence of tasks. Coordination of sequences of computational tasks in other contexts, such as the deployment of AI services, is done using workflow managers. A digital laboratory, whether fully automated or a VL, needs such functionality as well.
        
        The simplest workflow managers, such as Snakemake and Nextflow let users specify pipelines of steps that can run in parallel and track all inputs and outputs \cite{molder2021sustainable,di2017nextflow}. Kedro organizes machine‐learning projects into modular components with configuration management and automated testing support \cite{Alam_Kedro_2024}. Labber from Keysight handles low-level control of instruments and logs data in real time, though it defers decision making and step sequencing to user‐written code \cite{labber_keysight}. These tools support predefined or branched pipelines, but they cannot repeat tasks based on intermediate results or pause mid‐run for human input. Implementing conditional loops or interactive checkpoints still requires custom scripts or external controllers.

        In recent years, a parallel stream of tools similar in functionality has been developed specifically for workflows involving LLMs and other AI services.
        AI-agent frameworks package LLM calls and other AI services into agents that plan and execute multi‐step operations. LangChain offers prompt templates, call chaining, and simple memory for conversational use cases \cite{Chase_LangChain_2022}, and LlamaIndex supports indexing and retrieval of document collections \cite{Liu_LlamaIndex_2022}. AutoGPT and BabyAGI demonstrate agents that decompose goals, fetch web data, and invoke tools with minimal prompting \cite{Significant_Gravitas_AutoGPT}, while Microsoft's Semantic Kernel exposes application functions as AI‐callable plugins \cite{Microsoft_Semantickernel_2025}. While these frameworks simplify the development of, e.g., chatbots and document search substantially, they do not include built‐in support for running simulations, coordinating instruments, or defining structured experimental protocols. Despite these limitations, there have been recent attempts at building AI-assisted research tools using LLM workflows. For instance, \cite{swanson2024virtual} considered a system where a primary LLM manages the entire research cycle assembled from specialist agents (a chemist, a structural biologist, and a critic) in a VL, but embedding other kinds of tools besides LLMs would remain difficult.
        
        In summary, neither workflow managers nor AI–agent frameworks offer a unified environment integrating digital simulations, instrument control, iterative loops, and human‐in‐the‐loop interaction. Workflow managers lack native mechanisms for conditional repetition and interactive decision points, and agent frameworks focus on unstructured language tasks without primitives for the integration of hardware or simulations.
        
\section{AI-assisted Laboratories}
\label{sec:VAIlab}

    \subsection{Scientific Goals and the VL Concept}

        By providing a unified software tool for building a VL, we seek to support the broad application of AI techniques and AI-assisted research across scientific domains.
        The key aspects and advantages of this platform include:
        \begin{enumerate}
            \item \textbf{Reducing Overlapping Work:} By adopting a standardized approach, VLs minimize redundancy and overlapping efforts across different scientific disciplines. Traditional platforms often focus on specific areas, leading to isolated solutions that cannot be easily adapted or extended to other fields. 
            \item \textbf{Engaging the AI Community:} A unified formulation is particularly critical for engaging AI researchers. It allows them to apply their expertise to a wide array of scientific challenges without needing to tailor their approaches to isolated systems. This leads to innovation and cross-disciplinary collaboration, enabling AI developers to contribute more effectively to scientific advancements.
            \item \textbf{Transformative Advantages:} With more efficient and AI-assisted experimentation, researchers can explore new methodologies, optimize processes, and achieve insights that were previously unattainable with conventional techniques.
        \end{enumerate}

    \subsection{Overview}
        We introduce VAILabs as a concrete proof-of-concept open platform for running VLs. It allows for the description of a broad range of possible scientific discovery processes in a unified language and can be used for instantiating VLs in different scientific disciplines.  
        We follow the general recommendations for VL software outlined in \cite{klami2024virtual}; it is open source, uses a modular architecture where the individual tools required during the process are independent modules and not part of the core library, and is designed to be highly interoperable, allowing for seamless integration with existing data sources and computational tools.
        The library itself, as well as example workflows implementing real research tasks discussed later in Section~\ref{sec:usecases} are available at \url{https://anonymous.4open.science/r/VAI-Lab-7178}.

        \subsection{Audience}
        
        VAILabs is targeted at two audiences: (a) Domain scientists working on digitizing their laboratory research, and (b) AI researchers looking for opportunities to contribute to diverse scientific domains. The key functionalities and incentives for both audiences are summarized in Table \ref{tab:Benefits}, partly following the broader discussion in \cite{klami2024virtual}.

        \begin{table*}[ht!]
        \centering
        \caption{Complementary benefits of the VL platform for domain scientists and AI developers, and researchers.}
        \label{tab:Benefits}
        \begin{tabular}{cp{4.1cm}p{4.1cm}}
            \toprule
            \textbf{Key Benefits} & \centering \textbf{For Domain Scientists} & {\centering \textbf{For AI Developers and Researchers}} \\ \midrule
            \textbf{Tool Functionality} & Lightweight tool for lab operations with a low-entry barrier GUI and automatic data handling. & Provides open challenges and systematized research patterns for AI research. \\ \midrule
            \textbf{Module Flexibility} & Incorporates existing tools/devices as modules and facilitates switching between digital twin models and physical devices. & Facilitates problem structuring in a way that benefits all collaborators. \\ \midrule
            \textbf{Adaptability} & Enables easy integration of new modules, written in Python, without workflow recompilation. & Enables testing and demonstration of new methods in varied contexts. \\ \midrule
            \textbf{Systematic Language} & Provides a structured description of research processes with future potential for automated notes. & Allows research on decision-making and human-AI collaboration in complex tasks. \\ \midrule
            \textbf{Control and Interaction} & Users control interactions with domain-specific tools, retaining usual workflows. & Expanding user base for newly developed AI methods. \\ \midrule
        \textbf{Collaboration} & Allows exposing real research patterns and challenges to AI developers with minimal translation cost. & Facilitates human-AI collaboration and fosters consensus across diverse domains. \\ \bottomrule
        \end{tabular}
        \end{table*}

        For a \emph{domain scientist}, VAILabs offers a lightweight and adaptable tool that simplifies laboratory operations without disrupting existing workflows. Scientists deploy VAILabs locally, integrate their own measurement devices or computational tools as modules, and describe their experiments using a structured XML format. VAILabs supports complex research patterns, including iterative loops such as Bayesian optimization or reinforcement learning, which are difficult to express with traditional workflows. Built-in logging ensures reproducibility, while manual interactions and adaptive decisions are transparently recorded throughout the process.
        An integrated control and tracking system logs every step of the experiment, providing a transparent record of the workflow’s evolution. This not only ensures full reproducibility but also enables adaptive experimentation, where decisions made during execution are captured and documented as part of the scientific process.
        
                    
        An \emph{AI developer} would use VAILabs as a means of getting access to data and problem definitions that are relevant to scientific discovery processes in diverse fields. They contribute to the development of VAILabs by providing modules, ideally for tasks that reoccur in several domains, such as digital twin maintenance, experimental design, cognitive models of human decision-making, etc. 
        Importantly, for studying interactive AI assistance, a major challenge preventing efficient AI development today is the lack of relevant data and testing facilities. Widespread deployment of VL environments would largely resolve this problem, while also offering the AI researchers means for demonstrating their methods across multiple scientific domains.
        
    \subsection{Software Architecture}

\begin{figure}[htp]
            \centering
            \includegraphics[width=0.9\linewidth]{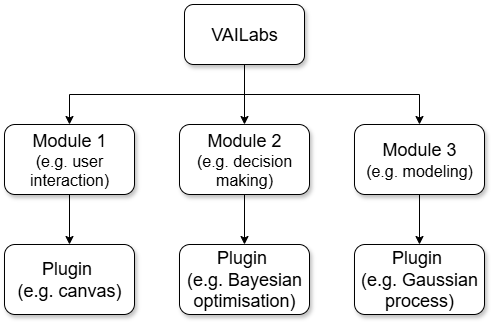}
            \caption{Proposed architecture. Each module within the framework requires a specific plugin that follows a module-specific standardized structure.}
            \label{fig:architecture}
        \end{figure}

        The VAILabs codebase is structured around a modular architecture (Figure \ref{fig:architecture}), where each module represents a distinct data processing or manipulation task. Within each module, various plugins offer specific methods or implementations to execute these tasks. For instance, the \emph{DataProcessing} module is responsible for data manipulation, and the choice of the plugin within this module determines the precise data processing method applied, such as the \emph{Binarizer} plugin, which converts data values to 0 or 1.
        
        While a plugin specifies the implementation details for a specific process, the core of VAILabs manages the setup and execution of these plugins. Each module contains a \emph{Core}, defining the required methods and attributes for a compatible plugin, as well as the instantiation and execution of the plugin. For example, in the \emph{DataProcessing} core, we define that a plugin like \emph{Binarizer} must implement methods such as \emph{transform()}, which requires an input dataset \emph{X} and optionally takes a \emph{copy} parameter to determine whether to return a new instance or modify the input data in place.
        
        In essence, the VAILabs software comprises three key components:
        \begin{enumerate}
            \item \textbf{Modules}: These are containers representing generic data manipulation or generation processes, with plugins populating them.
            \item \textbf{Plugins}: Specific implementations for running processes within modules. These may run multiple methods within.
            \item \textbf{Cores}: Responsible for managing background processes and handling plugins.
        \end{enumerate}
        
        This modular approach allows the flexibility to propose models compatible with diverse domains, as common techniques can be interchanged. One of the primary objectives of our research is to identify the essential modules required for VLs that can cater to various research projects.

        VAILabs is structured around a set of fundamental modules that define and execute the research workflow. At the foundation are two essential components: the \emph{Initialiser} and \emph{Output} modules. These modules act as the entry and exit points of the workflow, setting the scope and context of the scientific study.
        
        The \emph{Initialiser} module serves as the first step, establishing the problem domain. It contextualizes the problem and prepares subsequent modules with compatible plugins. The user specifies the data for the problem, and the framework automatically determines whether the problem is supervised or unsupervised, and the presence of data partitions.
        
        Simultaneously, the researcher setting up the pipeline outlines the desired problem, specifying how the modules should interact. This involves selecting required modules, detailing the pipeline, and, if necessary, defining loop iterations over the modules.

        The \emph{Output} module determines the data to be saved and its destination after executing the pipeline. It stores outputs from various modules at the specified location.
        
        With the input and output of the model defined, the domain, context, and research requirements are established. Various modules can then be employed to manipulate and analyze data towards a solution. We have identified the following modules as fundamental requirements for a fully functional framework:
        \begin{enumerate}
            \item \emph{Data Processing}: Responsible for data preprocessing tasks such as scaling, feature selection, or multivariate analysis. 
            \item \emph{Decision Making}: Utilizes techniques such as Bayesian Optimization (BO) to provide suggestions for the next experimental iteration. 
            \item \emph{Environment}: Creates and manages simulated or real-world scenarios in which the model operates. This includes tools and libraries like \emph{PyBullet} \cite{coumans2020} for physical simulations, or custom environments tailored to specific research needs. It also encompasses the integration of physical measurement devices, allowing real-time data collection and interaction with physical experiments.
            \item \emph{Modeling}: Captures information within the data, commonly used for data analysis, intermediate processing, or making predictions using classical supervised and unsupervised ML models.
            \item \emph{User Interaction}: Provides a means for user feedback, modification, or data analysis at specific points using a Graphical User Interface (GUI), e.g., altering BO suggestions based on expert knowledge.
        \end{enumerate}
        Each module comprises a set of plugins that correspond to the same modular family. After defining the data and pipeline, the researcher selects the plugins for each module. The framework adapts the available plugin options based on the problem domain. For example, an AI algorithm learning a user model would be implemented as a module focusing on adaptive learning algorithms that iteratively update and refine the user model based on interactions and feedback. Within each plugin, the researcher can choose the methods to execute, their order, and the parameters. Additionally, the user defines the data flow, specifying the input data for each module, required plugin functions, and parameters. Once the plugin specifications are complete, the final user can initiate the pipeline execution.

    \subsection{Interactive Experiment Control}

        One of VAILabs' most distinctive features is its support for real-time intervention during experiment execution, a capability lacking from standard workflow managers. Researchers can pause the pipeline at any stage to inspect intermediate results, validate progress, and adjust plugin configurations. This pause-and-check capability transforms experiments from rigid pipelines into interactive, adaptive processes.
        Such iterative interaction empowers researchers to fine-tune their workflows as they proceed, ensuring alignment with experimental goals and enabling informed decision-making at every step. Crucially, these interventions do not require restarting the pipeline, saving both time and resources.
        
        To support this flexibility, VAILabs provides a visual interface that clearly represents the stages of the experimental pipeline and their current status. Researchers can monitor execution in real time and easily navigate complex workflows. Moreover, plugin settings can be modified mid-pipeline, allowing dynamic reconfiguration based on insights from ongoing results.
        To avoid unnecessary interruptions, the system automatically proceeds to the next stage when no user input is detected. This balance between manual control and automation ensures efficient yet flexible experimentation.

    \subsection{Data Passing}
\begin{figure*}[!t]
        \centering
        \includegraphics[width=0.9\linewidth, trim=0.5cm 0cm 0cm 1cm, page=2]{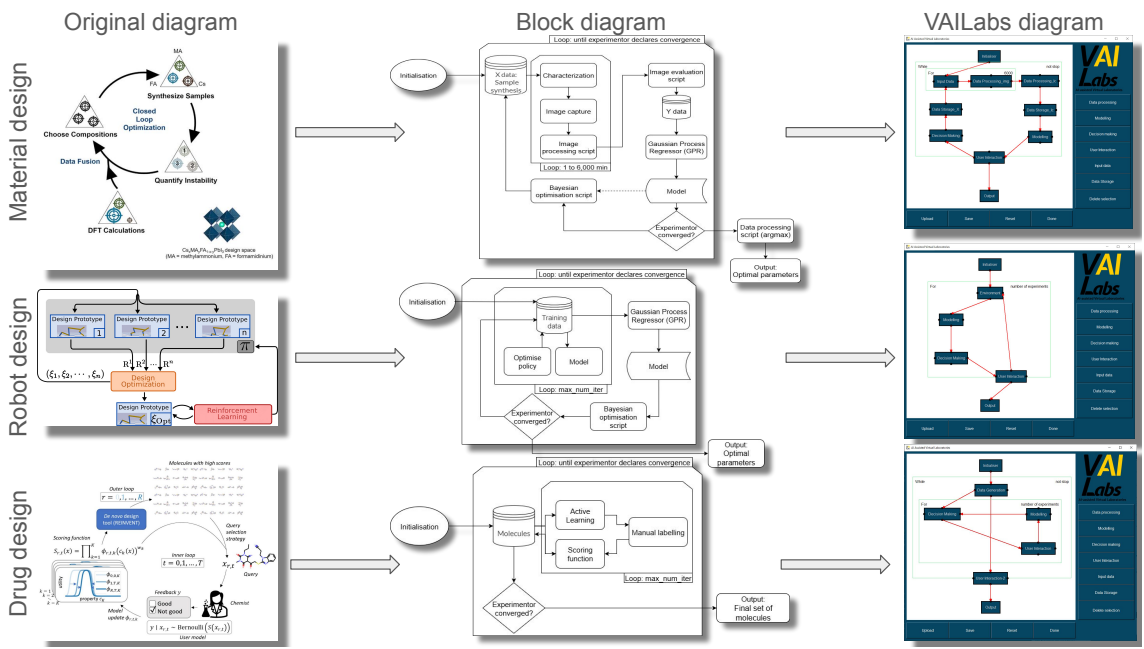}
        \caption{Mapping scientific workflows into the VAILabs platform. The first column shows original diagrams from three use cases: materials design (top), robot design (middle), and drug design (bottom). These conceptual workflows are redesigned as modular block diagrams (second column), highlighting their structure and data flow. The third column demonstrates how these workflows are implemented within the VAILabs platform, where each component maps to a functional module in a virtual laboratory workflow.}
        \label{fig:use}
    \end{figure*}
        In the realm of VLs, efficient data passing plays a pivotal role in ensuring the smooth flow of experimentation. This subsection delves into the concept of data passing, the alternatives explored, and our proposed approach for handling data within the VL environment.
        Data passing refers to the process of transferring information or data between different modules or components within a VL. It is a critical aspect of ensuring that each module can access the necessary input data and communicate its output to other modules seamlessly \cite{russell2005workflow}.

        We introduce a structured approach to data passing that enhances the clarity and manageability of the experimentation process. Each module within the VL defines the specific data it requires as input, a practice that not only clarifies dependencies but also ensures that the necessary information is readily accessible. This approach provides a clear framework for managing data and ensures that each module knows precisely what it needs to function effectively.
        
        To facilitate data access, modules can obtain their input information through two distinct mechanisms. First, they can load pre-existing data from a designated folder, allowing for the reuse of previously generated or external datasets. Alternatively, modules can use the output of previously executed modules within the same workflow. This dual approach grants flexibility to researchers, allowing them to access data from various sources and adapt their experiments as needed.
        
        Furthermore, our framework includes a systematic approach to the storage of execution output. We maintain a comprehensive record of the output generated by each module during execution, encompassing both final results and intermediate data. This record ensures transparency and traceability, enabling researchers to retrace the steps of their experiments and validate their findings.
        In addition, researchers have the flexibility to save the values or data they require for further analysis or reference. This user-defined data-saving feature allows researchers to tailor their data management to their specific needs and preferences, allowing for the seamless integration of experimental results into their research workflows. 
        
    \subsection{Foundation of VAILabs: XML-Based Workflow Description}
    
        
        At the heart of VAILabs lies an XML-based workflow description, representing the cornerstone of our proposal. This XML file serves as the essential tool for orchestrating research experiments, encompassing the entire research pipeline. It acts as a comprehensive blueprint, guiding researchers through the sequence of processes, the specific modules, data dependencies, and execution logic that define their experiments.
        
        The simplicity and user-friendliness of our XML template are paramount. We have designed it to be accessible and intuitive, ensuring that researchers, regardless of their technical background, can easily define their research workflows. This ease of use empowers both programmers and non-programmers, making the power of VAILabs accessible to a wide range of researchers.
        
        Flexibility and customization are key features of our XML-based system. Researchers have the freedom to fully customize their workflows, tailoring each process to their specific research needs. This adaptability enables researchers to fine-tune their experiments and respond to evolving research demands. You can find examples of the XML workflow in the repository.
        
        

\section{Demonstrations in Scientific Problems}
\label{sec:usecases}



    Our main goal is to enable running different kinds of research activities using the same tools. To demonstrate having achieved this, we show how three different kinds of research processes (Figure~\ref{fig:use}) can be re-implemented using VAILabs. Each example corresponds to a real published research activity that was originally carried out without VAILabs. For each case, we provide a brief executive summary and a compressed explanation of the original research workflow. We then map the workflow to a VL description and re-implement the process in VAILabs, followed by a discussion on how the original research would have benefited from VAILabs had it been available then.
    
    One example consists of traditional empirical science, where physical experimentation is carried out and combined with ML approaches to optimize the experimentation on different material properties. The other example is purely digital and requires a simulation environment where robot design and behavior co-adaptation can be combined without physical prototypes. Finally, the third example combines ML models (for molecule generation) with continuous human interaction, as an example of a typical human-in-the-loop workflow.

    
    \subsection{Case A: Material Design}

    \paragraph{Overview:} The task is to design improved solar cell materials. Different perovskite composites are evaluated using empirical experiments, and a Bayesian optimization algorithm is used for iteratively recommending the composites. The original work of Sun et al. \cite{sun2021data} used a field-specific BO library and discovered an improved, more stable composition despite covering less than $2\%$ of the discretized space of possible compositions. The process can be easily mapped as a VL, with some savings in implementation effort, and for easier identification of opportunities for further AI support.
        
        \paragraph{Original research:}
        New materials are needed in the development of various new materials-intensive technologies, such as new types of solar cells, and to speed up the experimentation, BO tools are increasingly being used \cite{jin2023bayesian}, often using field-specific libraries like BOSS \cite{moss2020boss}. 
        Sun et al. \cite{sun2021data} screened new perovskite compositions to be used in perovskite solar cells, to find environmentally stable materials that can survive use in extreme weather conditions. 
        Certain perovskite compositions can be synthesized relatively easily, and their stability can be evaluated using high-throughput degradation test chambers where the samples       
        are exposed to heat, humidity, and illumination stress. Due to the slow testing time (100 hours), the number of considered compositions should be minimized, even though the testing is largely automated (e.g., the photographing the samples with 5-minute intervals).

        The research as described by Sun et al. \cite{sun2021data} combines BO for recommending which perovskite compositions to synthesize and test. The target function for BO is a cheap proxy variable for the stability of the samples, instability index $I_c$ computed through the analysis of color changes in the sample. The iterative BO process continues until the researcher declares the solution to be sufficiently stable.
        With the described approach, Sun et al. \cite{sun2021data} were able to determine the environmentally stable composition region among \(Cs_xMA_yFA_{1-x-y}PbI_3, x, y \in [0,1]\) (MA, methylammonium; FA, formamidinium) perovskites by sampling only $1.8\%$ of the discretized composition space (112 samples and four degradation tests in total), which is a clear improvement compared to more traditional Edisonian or one-variable-at-a-time optimization approaches. The collected data, synthesis results, analysis codes, and the BO algorithm modified for this experiment are shared openly online.
        
        \paragraph{VL workflow:}
        
            \begin{figure}[htbp]
                \centering
                \begin{subfigure}[b]{0.49\textwidth}
                    \centering
                    \includegraphics[width=0.9\textwidth,keepaspectratio]{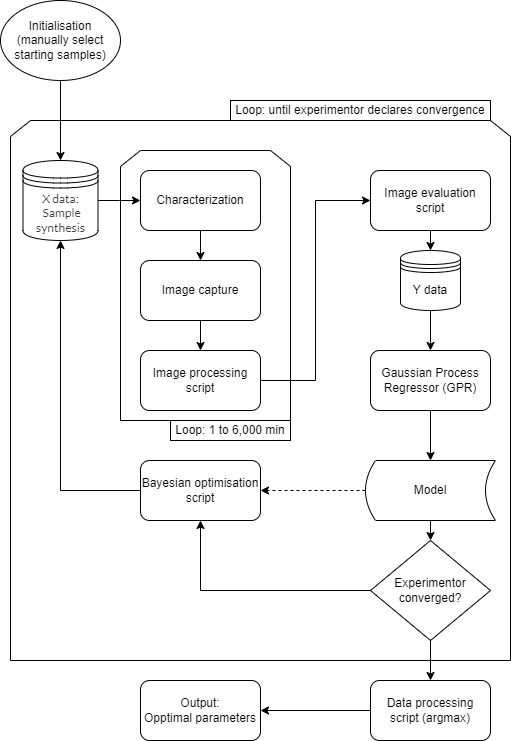}
                    \caption{Block diagram}
                    
                    \label{fig:BD_crystal}
                \end{subfigure}
                \hfill
                \begin{subfigure}[b]{0.49\textwidth}
                    \centering
                    \includegraphics[width=0.9\textwidth,keepaspectratio]{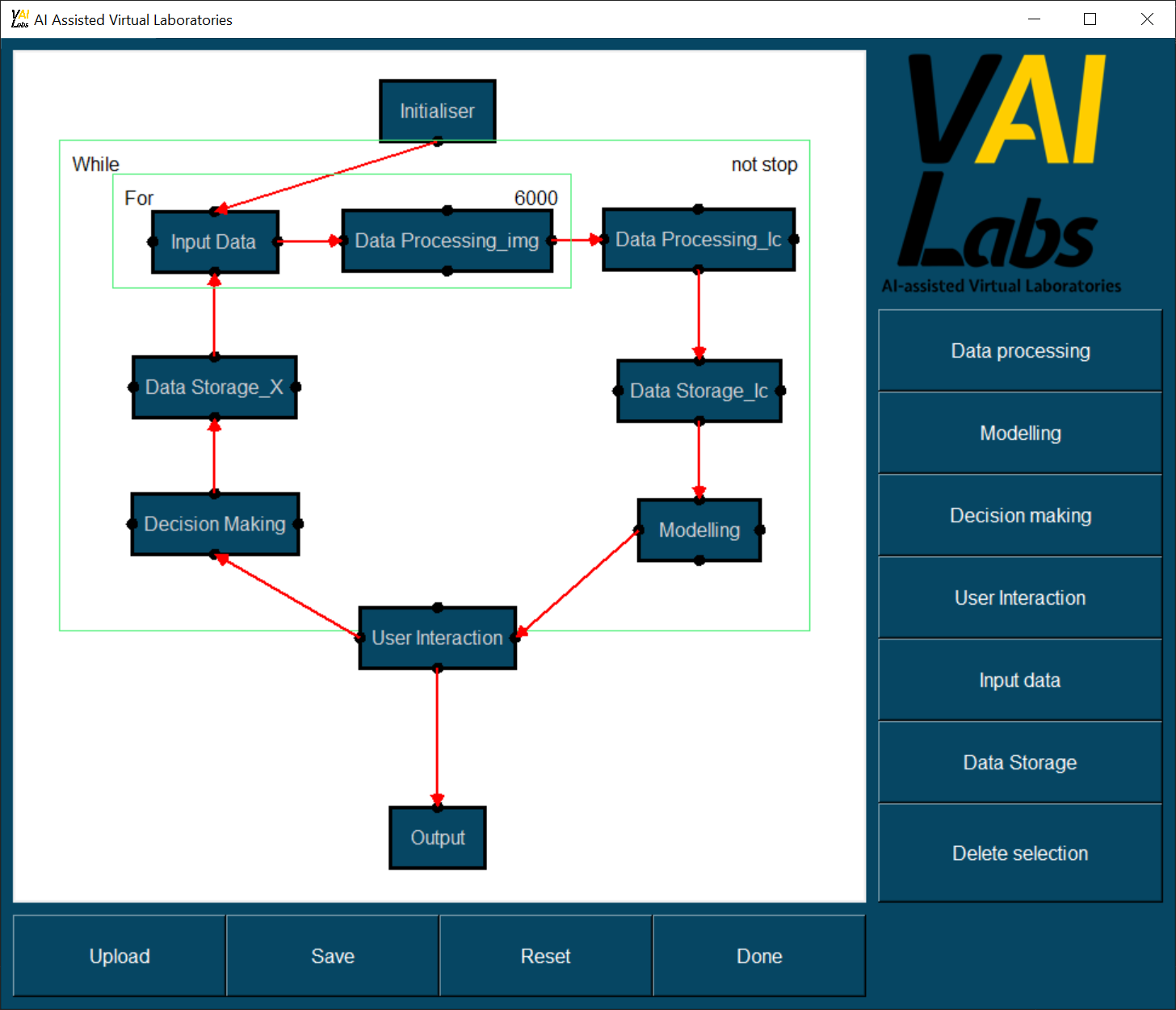}
                    \caption{VAILabs pipeline}
                    \label{fig:VP_crystal}
                \end{subfigure}
                \caption{Workflow representation for the material design problem as discussed by Sun et al. \cite{sun2021data}.}
                \label{fig:crystalDesign_blocks}
            \end{figure}

        This is a prototypical research workflow in material screening, but also in numerous other empirical sciences. At the core is a loop where different alternatives are sequentially considered, and all evaluations are done with physical experimentation that here includes both synthesis and evaluation of the samples. However, the research practice already uses digital tools: The recommendations are made by a BO algorithm, the testing chamber is automated, and the results are analyzed by a computational algorithm processing the images captured during testing. Finally, the researcher makes an explicit decision of termination, rather than running the iteration for a pre-determined number of rounds.


            Figure~\ref{fig:crystalDesign_blocks} describes the process as a VL workflow, isolating the individual components, and shows what it looks like in the VAILabs format, where specific choices of the methods are represented as abstract modules. The automated imaging during the testing chamber becomes a loop, the specific models (GPR, image processing) are general \emph{Modelling} modules, \emph{Data Storage\_Ic} and \emph{Data Processing\_IC} modules are used for transferring data, and the termination is implemented as \emph{User Interaction}.

    \paragraph{Effort and benefits:} In this example, the image analysis technique is specific for the domain and closely linked to the specific research technique. This element would remain as before and be interfaced from VL, with the data passing implemented by the researcher. The BO algorithm, however, is general, and in fact, an external BO library was already used in the original work. From the perspective of algorithm development, there would hence be no notable difference between VAILab and the original work.

    However, there would still have been clear practical benefits for the researcher. VAILab takes care of the internal data passing and storage in a systematic fashion; the researcher would not be limited to using a specific BO library, and the process itself would be more systematic and easier to communicate for outsiders. Finally, the explicit mapping would make the seemingly trivial choices more explicit, with opportunities for improvement. For instance, a study of different termination criteria is clearly out of scope for a material scientist, but if, e.g., an algorithm that recommends a few extra experiments to validate the decision was available in the VL environment, it could easily be tried out here.

    The original work was exemplary in releasing both the implementation and results openly. VL would make this easier for others, supporting good practice, but it would also have been useful in this case. As of now, the data release remains isolated, and unfortunately, it is unlikely that a BO researcher would download the original data release. However, had the work been released as an open VL instance, any researcher providing a new BO algorithm for VLs in general would be likely to evaluate it also on this data.

    \subsection{Case B: Co-adaptation of Robot Design and Behavior}    

    \paragraph{Overview:}
    Luck et al. \cite{luck2020data} introduce a new principle for designing robots, by jointly optimizing the physical design of a robot and learning how to best behave in the environment, aiming for a design for which, e.g., learning safe interactions is easier. This co-design is done using reinforcement learning and black-box optimization (e.g., BO) in simulation environments to identify good physical designs before manufacturing them. As a purely digital and automated research phase, it maps naturally as a VL.

    \paragraph{Original research:}
        Robots have been largely designed manually by human engineers and have been optimized for controllability and ease of mathematical modeling of movements. Hence, today's robots are dominantly made of rigid material and simple shapes, and are not necessarily optimal, e.g., for energy efficiency or safe interaction with the environment. Instead of designing the algorithms for controlling the robots conditional on the existing design, \emph{co-adapting} design and behavior at the same time allows one to uncover better optimized design-behavior combinations. 

        Luck et al. \cite{luck2020data} proposed a solution where the structure of the robot and the behavioral policies are simultaneously optimized in a simulated environment. This is essentially a bi-level
        optimization problem: On the lower level, given a robot morphology, one aims to learn and optimize movement policies given data collected on the current robot using reinforcement learning. On the higher level, given the final learnt policy and the history of considered robot morphologies, we can predict how to vary design parameters to optimize and uncover new, better optimized robot bodies.  The prediction of future performance of robot designs can be realized by optimizing in-silico robots directly in simulation with, for example, BO or evolutionary methods. One can also directly model the lifetime performance of robots with the episodic return (i.e., Q-Value networks) and deep neural networks as a surrogate function, which can then be queried for predictions of robot design performances.

\begin{figure}[htbp]
            \centering
            \begin{subfigure}[b]{0.49\textwidth}
                \centering
                \includegraphics[width=0.9\linewidth]{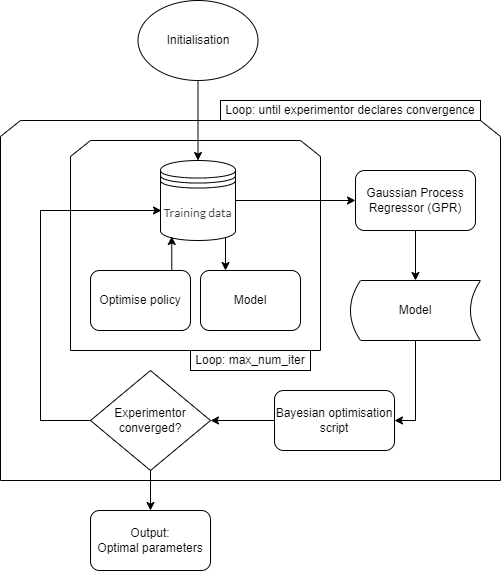}
                \caption{Block diagram}
            \end{subfigure}
            \begin{subfigure}[b]{0.49\textwidth}
                \centering
                \includegraphics[width=0.995\textwidth]{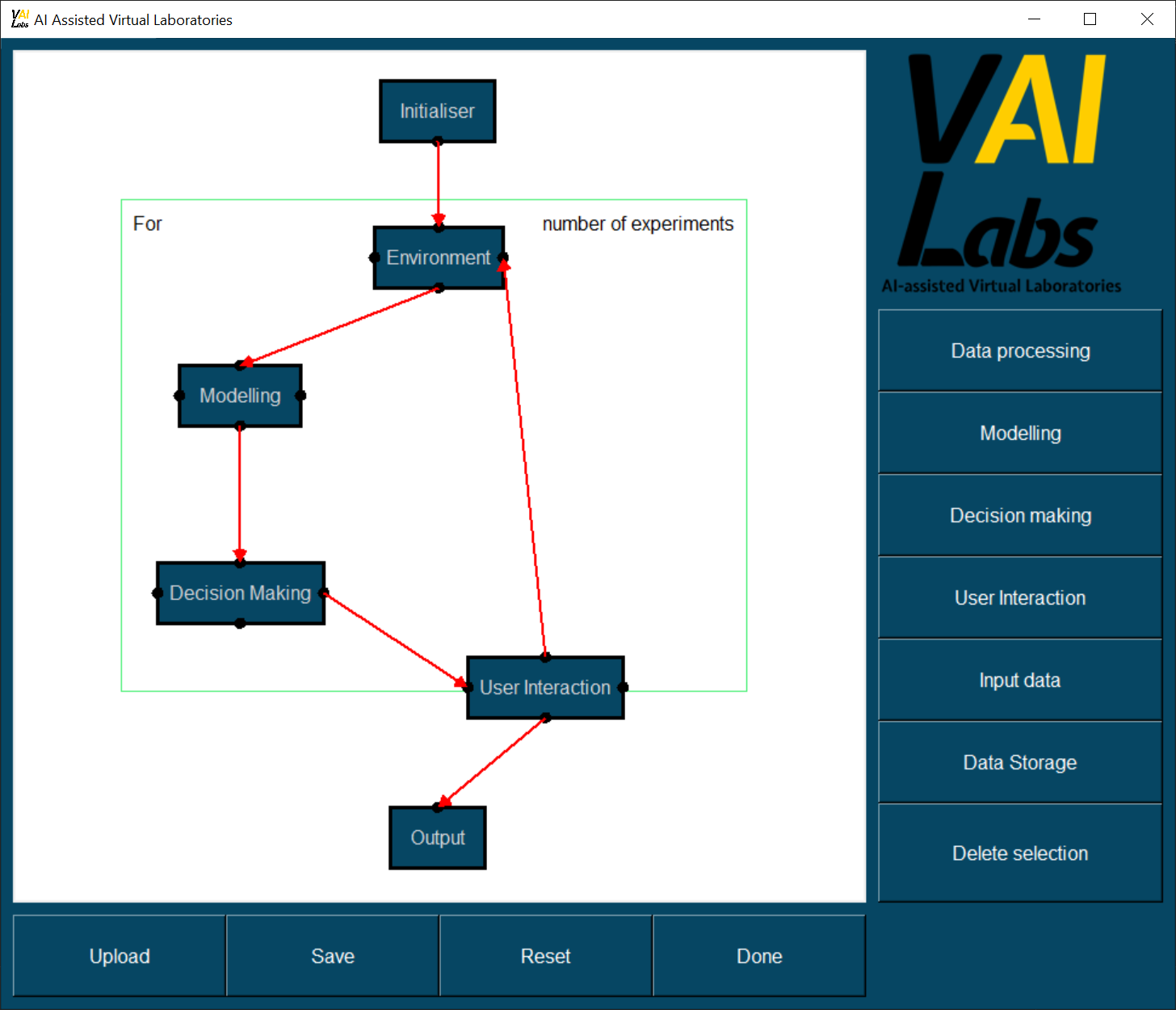}
                \caption{VAILabs pipeline}
                \label{fig:VP_robotics}
            \end{subfigure}
            \caption{Co-adapting the robot behavior and design of robots as discussed by Luck et al. \cite{luck2020data}. }
            \label{fig:robotics_design}
        \end{figure}

        \paragraph{VL workflow:}
        Figure \ref{fig:robotics_design} maps the co-adaptation problem to VAILabs, as two nested loops:        
        (1) the inner behavior-optimizing loop and (2) the outer performance-optimizing loop. Both could be implemented using various algorithms. Here, the inner loop uses model-free reinforcement learning and collects the acquired data in a local replay buffer.
        This training loop is fully embedded in the Environment module in Figure \ref{fig:VP_robotics} and does not require interaction with a human experimenter. 
        
        The results of the inner loop are handed to the outer optimization loop. It learns the mapping between robot design parameters and performance metrics, such as rewards, costs, and episodic return. This is implemented using BO. 
        Once a predictor or surrogate function is learnt, a selection of possible design candidates may be presented to the user, who can fine-tune the selected design parameters (\emph{User Interaction}) before entering the inner loop again.
        Once this process has been repeated several times, the final optimized design can be constructed physically.

    \paragraph{Effort and benefits:}
    Implementing the research process in VAILabs requires effectively only the description of the two nested optimization processes. The algorithms themselves, as well as the reinforcement learning environment, are standard elements that are used within the VL.

    The key benefit of mapping this fully automated pipeline to a VL is that it reveals opportunities for researcher interaction. For instance, a human expert could filter the recommendations of the outer optimization loop to prune out uninteresting morphological candidates before running testing (similar to what is done in Case C) or the fixed number of inner loop iterations could be replaced with researcher-guided termination criteria (similar to what is done in Case A). Without the VL formulation, such extensions would need to be developed case by case, but VL makes it easy to add such interaction patterns following previous examples in other workflows.

    \subsection{Case C: De Novo Drug Design}

    \paragraph{Overview:}    
    Generative models that can synthesize new molecule structures \cite{tang2024survey} are increasingly being used as a part of the drug design process \cite{pang2024deep}, as digital components of a laboratory. Helping researchers better use such tools and hence increasing their value is the main goal of the VL concept. Here we map to the VAILabs an example study where preference feedback provided by a researcher is used to steer the generative model \cite{sundin2022human}, as an example of human-AI collaboration that generalizes also to other research tasks involving generative AI components.

        \paragraph{Original research:}
            In drug design, chemists aim to discover molecules that have specific properties, which could lead to new cures for diseases. The search space for these molecules is vast and challenging for chemists to navigate. Reinforcement learning and software like REINVENT \cite{olivecrona2017molecular} address this issue by balancing exploration and exploitation to efficiently search the molecular space. From another angle, it is not always easy to come up with a reward function for the reinforcement learner to provide us with decent molecules. On the other hand, chemists have a good intuition as to what a good molecule is, but cannot define it mathematically. Therefore, it would be advantageous to acquire a reward or scoring function through engagement with a chemist. The first work to do so is by Sundin et al. \cite{sundin2022human}, who used chemist feedback and inferred the parameters of the scoring function using data collected from the chemist. Here, we will consider the simulated human feedback case. 
            
        \paragraph{VL workflow:}
            \begin{figure}[htbp]
                \centering
                \begin{subfigure}[b]{0.49\textwidth}
                    \centering
                    \includegraphics[width=0.85\linewidth]{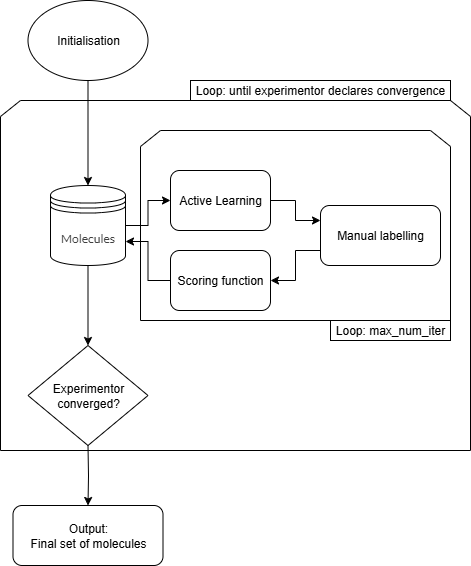}
                    \caption{Block diagram}
                    \label{fig:BD_drug}
                \end{subfigure}
                \hfill
                \begin{subfigure}[b]{0.49\textwidth}
                    \centering
                    \includegraphics[width=\textwidth]{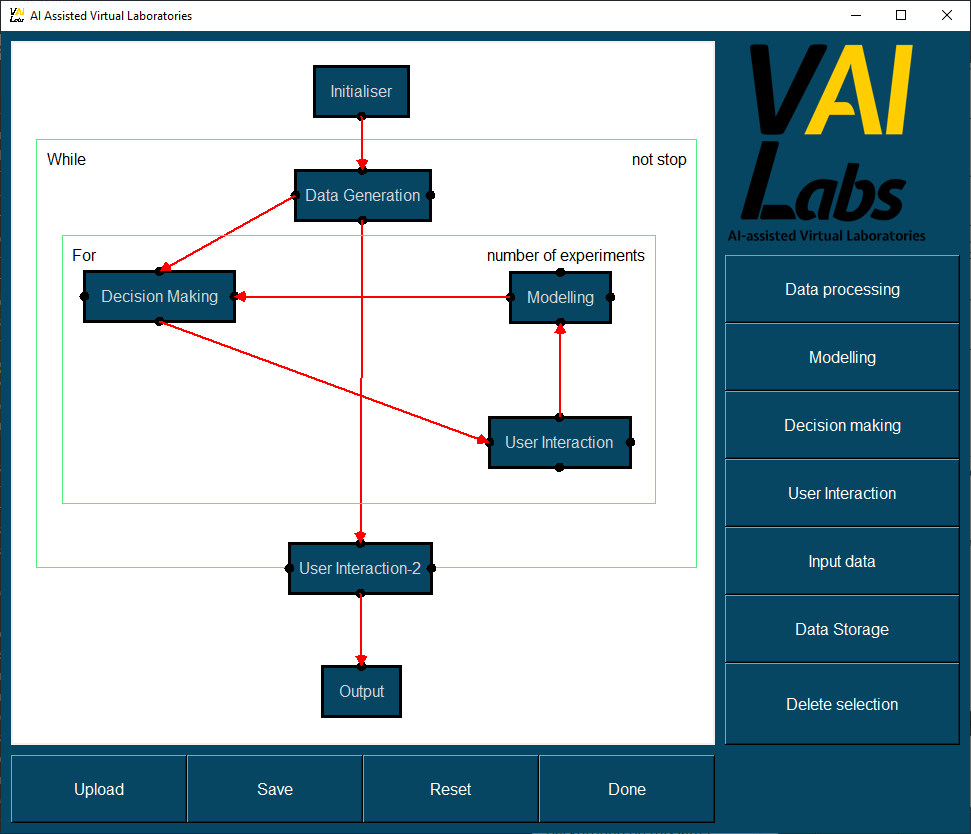}
                    \caption{VAILabs pipeline}
                    \label{fig:VP_drug}
                \end{subfigure}
                \caption{Human-in-the-loop assisted de novo molecular design workflow as discussed by Sundin et al. \cite{sundin2022human}}
            \end{figure}
            
            In this section, we discuss the mapping of Task 1 of Sundin et al. \cite{sundin2022human} to VAILabs.
            To model the human-in-the-loop de novo design using VAILabs, there are a number of components that are used:
            \begin{itemize}
                \item \textbf{Molecule generator}: which is a decision-making module that, given a scoring function as input, generates a set of molecules. To do so, the molecule generator calls REINVENT~\cite{blaschke2020reinvent}.
                \item \textbf{Active learning selection strategy}: which is a decision-making component that, given a scoring function and the set of molecules generated by the molecule generator, produces one molecule. This is a mechanism to select the most informative molecule to be labeled by the chemist. 
                \item \textbf{Human feedback}: which is a user-interaction module that, given a molecule as input, labels it as good or bad in the form of a 0-1 binary feedback. The labeling can be performed either by a real chemist or using a "synthetic chemist" to obtain the label as described by Sundin et al. \cite{sundin2022human}. 
                \item \textbf{Scoring function inference}: which is a modeling module that, given the set of labeled molecules as input, infers the parameters of the scoring function and outputs a scoring function.
            \end{itemize}
        The workflow, as can be seen in Figure~\ref{fig:BD_drug}, includes two main loops: The outer loop is responsible for generating a set of good molecules using the ``Molecule generator'' and the inner loop is an active learning loop, which is responsible for selecting the most informative molecule at each iteration. The chosen active learning molecule is then sent to the chemist, who labels it, and the molecule-label pair is an input to the inference procedure, which estimates the parameters of the scoring function. The corresponding diagram of this workflow in VAILabs can be seen in Figure~\ref{fig:VP_drug}.

    \paragraph{Effort and benefits:}
    As in the previous example, the VL implementation is relatively straightforward, consisting of two nested loops and the interaction components. The overhead compared to implementing it as a stand-alone code is minimal.

    Even though the workflow was designed originally for a specific task in drug design and assumes the evaluation of the candidates is done by the researcher, it directly applies for any task where a generative AI component is used for proposing alternative candidates and a separate module validates their suitability. It only requires that the generative AI component allows some form of steering the outputs, which is the case for all conditional generators; the REINVENT generator used here assumes a reinforcement learning mechanism for that, but the inner loop could equally well be used to estimate direct covariates for the generative model. The validation, here done manually by the chemist, could be replaced by e.g. a trial experiment, a measurement device, or another computational model. This would only require implementing an alternative for the manual labeling node.

\section{Discussion}

    Scientific exploration is undergoing a transformation, driven by the increasing modularization and openness of research tools. This shift enables broader participation and fosters innovation by lowering technical barriers and facilitating collaboration across disciplines. VLs act as enablers in this evolution, spreading novel methods and promoting reuse, reproducibility, and adaptation across domains. By encouraging open-source implementations and standardized interfaces, VLs ensure that scientific practices evolve in step with emerging technologies while remaining accessible and verifiable.
    
    Our modular framework contributes to this vision by formalizing key components of experimental pipelines. For example, in our first demonstration using a standard BO algorithm, the modular architecture allows for straightforward substitution with more advanced variants (e.g., Monte Carlo Tree Search), enabling researchers to upgrade their optimization routines without modifying the broader workflow. Similarly, in our third example involving molecule generation, alternative generative models can be tested interchangeably within the same pipeline, accelerating development and benchmarking in AI-assisted chemical synthesis.
    
    The benefits of modularity extend beyond algorithmic flexibility. Innovations in human-computer interaction, such as evolving from binary feedback to more expressive interfaces, can be smoothly integrated. Even though in our examples we deliberately replicated the original workflows from prior studies, a researcher could trivially reuse the same BO-based decision-making module from the robotics experiment in the drug discovery setting. Similarly, the user-feedback interface from the materials science scenario could be inserted into any compatible pipeline. This plug-and-play capability encourages researchers to systematically assess the value of independently developed interaction modules in new contexts.
    
    Importantly, our framework complements the emerging trend of using LLMs as co-researchers or intelligent assistants. While recent work explores how LLMs can participate in the research process (e.g., guiding experiment design, generating hypotheses, or interpreting results), these systems currently lack structured integration into formalized experimental protocols. Our approach addresses this gap by focusing on the systematization of scientific workflows, providing a structured environment in which LLMs and other AI components could be reliably embedded, evaluated, and improved. This opens up the possibility of combining the free-form reasoning capabilities of LLMs with modular pipelines that enforce reproducibility and interpretability.
    
    Despite its strengths, our framework also highlights open challenges in building truly general-purpose AI-assisted research environments. Mapping arbitrary scientific tasks into a modular structure remains a work in progress, and many of the benefits of a unified software architecture would only be seen once the tools are broadly used. Our current prototypes demonstrate feasibility, but general usability across domains will require further development of shared standards, robust defaults, and interface conventions. The framework exposes many of the concrete difficulties that remain unsolved in the community, such as designing user-friendly tools for specifying workflows and ensuring semantic compatibility across modules.
    Nevertheless, this openness is a feature rather than a flaw: by making the environment publicly available and modular, we encourage AI researchers and domain scientists to engage with and extend the system. This transparency promotes the systematic development of interoperable tools, fosters reuse and benchmarking, and accelerates collective progress toward realizing the potential of Virtual Laboratories.

\section{Conclusions}

    The main goal of the paper was to provide a proof-of-concept implementation of a general software tool for facilitating the operation of Virtual Laboratories that seamlessly combine the digital and physical world and enable AI-assisted research. We explained why previous attempts, ranging from purely virtual education environments and completely automated robot scientists to workflow managers and LLM-driven research loops, are insufficient for supporting the domain-agnostic digitalization of scientific research. We then characterized a potential software tool that resolves some of the open aspects.
    By mapping three distinct kinds of real research workflows --- each varying in the degree of human interaction and digitalization --- to our proposed architecture, we showcased the framework's ability to unify and streamline scientific research.
    
    Our examples highlighted several key advantages:
    \begin{enumerate}
        \item \textbf{Standardized Optimization}: In the first example, we showed how a standard BO algorithm could be easily substituted with more advanced variants, enhancing the optimization process without necessitating significant changes to the existing workflow.
        \item \textbf{Flexibility in Component Integration}: The third example illustrated the ease with which alternative molecule generators could be incorporated into the framework. This capability extends to the human interaction component, where simplified binary feedback mechanisms could be upgraded to more sophisticated interfaces, such as eye tracking or brain-computer interfaces (BCIs), to improve user interaction.
        \item \textbf{Public Accessibility for AI Research}: By providing a publicly available environment, VAILabs enables AI researchers to study and experiment with these components independently. This promotes a collaborative research culture and accelerates the development of innovative solutions.
    \end{enumerate}
    
    These capabilities underscore the potential of VAILabs to re-run previously conducted scientific research processes using a common tool layer, supporting reproducibility and enabling independent AI research focusing on AI assistance. Many of the benefits of the tool, however, would require broad deployment of the framework across the sciences. This requires both further development of the concept and the software environment as a community effort, and this paper is to be seen as a first step demonstrating the feasibility of the vision.
    



\bibliography{bibliography}

\end{document}